\documentclass[12pt, aps, pra, showpacs, floatfix]{revtex4}
\usepackage{amsmath, amsthm, amssymb}
\usepackage{graphicx}
\usepackage[usenames, dvipsnames]{color}
\usepackage{hyperref}
\hypersetup{colorlinks=true, linkcolor = blue, citecolor = blue, urlcolor = blue}
\newcommand{\mathd}{\mathnormal{d}}
\newcommand{\mathe}{\mathnormal{e}}
\newcommand{\mathi}{\mathnormal{i}}
\newcommand{\mathhc}{\mathrm{H.c.}}
\newcommand{\mathcc}{\mathrm{c.c.}}
\newcommand{\opH}{\hat{\mathcal{H}}}
\newcommand{\nullket}{|0\rangle}

\begin{document}
\title{Photon distribution function for propagation of two-photon pulses in
waveguide-qubit systems}
\author{Oleksandr O. Chumak}
\email{chumak@iop.kiev.ua}
\author{Evgeniy V. Stolyarov}
\affiliation{Institute of Physics, National Academy of Sciences of Ukraine, Prospekt Nauki 46, 03028 Kyiv, Ukraine}
\begin{abstract}
 Propagation of a two-photon pulse in a waveguide coupled to a
two-level system (TLS) is studied. The pulse is formed by two
spatially separated identical wave packets. A set of equations
governing the dynamics of the photon distribution in the
configuration-momentum space is derived and solved. It is shown that
the distribution function can be negative that manifests its
quasiprobability nature. A spectrum of the reflected light is found
to be narrower than that of the transmitted light that features a
pronounced filtering effect. Average numbers of the transmitted and
reflected photons and their variances are shown to be dependent not
only on the pulse widths and the light-TLS interaction but also on
the pulse separation that can serve as an effective controlling
parameter. Our approach is generalized for the case of an $n$-photon
Fock state.
\end{abstract}
\setcounter{page}{1}
\maketitle
\section{Introduction} \label{sec:intro}
  The problem of interaction of few-photon pulses with two-level
systems (TLSs) attracts an increasing interest. First of all, it is
due to the development of quantum information processing (QIP)
devices. The TLSs are the simplest implementations of the stationary
quantum bits (qubits).  In practice, various multilevel systems are
used: trapped ions \cite{cir,mon} or neutral atoms \cite{deu},
superconducting Josephson junctions \cite{wal}, semiconductor
quantum dots \cite{ben} {\it etc}. Nevertheless, in many cases those
multilevel systems can be modeled as TLSs. This simplification is
quite reasonable if the frequency of the incident radiation is close
to the transition frequency between the corresponding pair of levels.

Recent experiments show that photons can act as transmitters of
quantum states between distant stationary  qubits \cite{mats,moe}.
Moreover, stationary qubits are able  to controllably generate
correlations between photons.  The qubits, connected by optical or
microwave waveguides, form scalable chip-based circuits \cite{blai}.
These circuits are considered now as a potential hardware basis for
the QIP systems.
Properties of photons propagating in waveguide-qubit systems
have attracted increasing interest.

Theoretical description of propagation of few-photon pulses
in waveguides coupled to a TLS can be found in numerous publications \cite{dom, koj, hof, yud, shen, sfan, zheng, fan, xufan, rep}.
For example, a ``collision" of two wavepackets at a TLS was studied \cite{dom}.
A striking difference in the interaction of the Fock state and
coherent state wavepackets of the same photon number
was illustrated. It was shown, that photon-TLS coupling induces
correlation between photons that can be interpreted as their ``interaction".
This controllable photon-photon interaction may be used for generation
of spatiotemporal entanglement and four-wave mixing effects \cite{koj, hof}.

Evolution of the photon-TLS was studied in the Heisenberg picture in
\cite{dom}. In contrast, the authors of \cite{koj, hof, yud, shen,
sfan, xufan, zheng} preferred the Schr\"{o}dinger picture. The
theoretical analysis is simplified if the incoming and outgoing
radiation fields are away from the TLS and, accordingly, are outside
of the interaction range. In those regions the evolution of fields
is as if there is no interaction with the TLS. In this situation the
initial and the final states of the radiation are connected by the
$S$ matrix whose elements can be extracted from the eigenstates of
the full interacting Hamiltonian. Provided that the $S$ matrix is
known, the outgoing field can be expressed via the entering field. A
rigorous program to construct the complete scattering matrix which
is applicable for two or more photons was developed in \cite{shen,
sfan, xufan, zheng}. Using that technique the physical quantities
like transmission or reflection coefficients can be obtained
analytically. The above approach was extended in Ref. \cite{zheng}
for coherent-state wavepackets with arbitrary photon numbers.

Further studies \cite{fan, rep} were based on the input-output formalism
of quantum optics \cite{vog}. One- and two-photon scattering with a
TLS was analyzed. The relationship between the input-output
operators, which are inherent for the Heisenberg picture of the
evolution equations, and the photon scattering matrix was derived.
It was shown that these approaches are equivalent. At the same time
the authors of \cite{fan} inferred that the input-output
approach was more elementary than the techniques developed earlier in
\cite{shen, sfan, yud}.

The Heisenberg picture is suitable for using the formalism of a
phase-space distribution function \cite{ber} which
provides a comprehensive description of the system. Within this
approach the evolution of wavepackets in the coordinate space as
well as in the momentum space can be analyzed in detail \cite{sto}.
The operator of the phase-space distribution function, $\hat{f}(x, p, t)$,
represents the photon density in the coordinate-momentum $(x, p)$ phase
space. It was shown in Ref. \cite{sto} that the average value of the
phase-space distribution function, $\langle\hat{f}(x, p, t)\rangle$,
may be negative at some domains of phase space which indicates that
$\langle\hat{f}(x, p, t)\rangle$ describes a quasiprobability rather
than the probability of the photon density in the phase space.

In this work, we extend our previous studies \cite{sto} to the case of few-photon Fock states of the ingoing field.
We consider dynamics in the phase space and statistical properties of two-photon pulses
whose initial state is represented by a sequence of two single-photon wavepackets.
The initial distance between them is a free parameter which controls the correlation of outgoing photons.
Then we outline a general scheme to study systems with an arbitrary number of
photons.

The paper is organized as follows. The model Hamiltonian and the
initial state are drawn in Sec. \ref{sec:sec_2}. In Sec.
\ref{sec:sec_3} the set of equations describing dynamics of the
system is derived and solved. Evolution of two-photon pulses in
the phase-space is studied. In Sec. \ref{sec:sec_4} the statistical
properties of the outgoing photons are investigated. The obtained results
are summarized in Sec. \ref{sec:summ}. Derivation of
useful operator relations used throughout the paper is delegated to
Appendix \ref{sec:app_a}. In Appendix \ref{sec:app_b} we demonstrate
a generalization of the method for the case of an $n$-photon Fock
state input.
\section{The Model} \label{sec:sec_2}
\subsection{The Hamiltonian of the model}
\begin{figure}[t]
  \centering
  \includegraphics[width=0.7\textwidth]{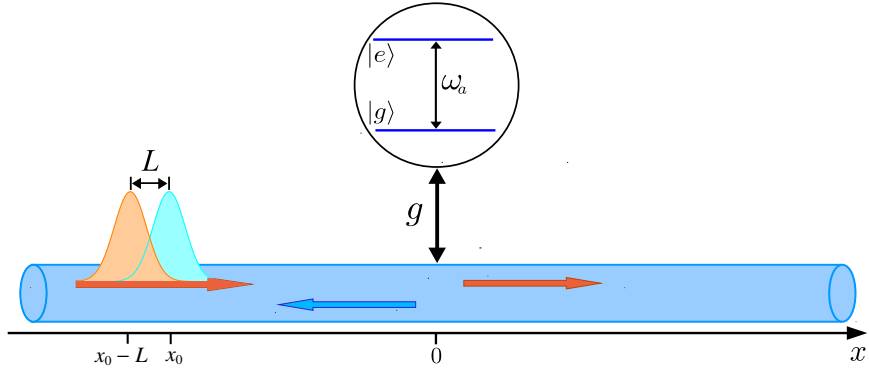}
  \caption{(Color online) Scheme of the system under consideration. A qubit modeled by a TLS is positioned at $x=0$ and coupled equally to waveguide modes propagating from the left to the right and vice verca.
  The ingoing two-photon state is represented by two single-photon pulses, which can be separated by distance $L$. The initial positions of the pulses are $x_0$ and $x_0-L$. \label{fig:fig_1}}
\end{figure}

The system we study consists of a TLS (qubit) coupled to photons propagating in both directions in a one-dimensional waveguide.
Figure \ref{fig:fig_1} displays the scheme of the model system.
Ground and excited states of the TLS are denoted  as $|g\rangle$ and $|e\rangle$, respectively.
The system is modeled by the Hamiltonian
\begin{equation} \label{eq:eq1}
  \opH = \opH_0 + \opH_{int}.
\end{equation}
Here $\opH_{0}$ describes the free evolution of a TLS and the field in the waveguide.
Assuming that the waveguide modes form a one-dimensional continuum \cite{shenf}, it is given by
\begin{equation} \label{eq:eq2}
  \opH_0 = \omega_{a} \, \sigma_{+} \, \sigma_{-} + \int \mathd p \,
  \left(\omega^{l}_{p} \, l^{\dag}_{p} \, l_{p} + \omega^{r}_{p} \, r^{\dag}_{p} \, r_{p}\right),
\end{equation}
where $\omega_{a}$ is the transition frequency, $\sigma_{+} =
|e\rangle\langle g|$ and $\sigma_{-}=|g\rangle\langle e|$ are
raising and lowering operators obeying the Pauli matrices algebra,
$l^{\dag}_{p}$($l_{p}$) and $r^{\dag}_{p}$($r_{p}$) are respectively bosonic creation
(annihilation) operators of the photons propagating in the waveguide
from the left to the right side and vice versa. In what follows we
use terms ``$l$-mode ($l$-photon)" and ``$r$-mode ($r$-photon)" to denote
the left-to-right and right-to-left propagating modes (photons),
respectively. Photon frequencies, $\omega^{l,r}_{p}$, linearized
with respect to momenta $p$ (see Ref. \cite{sf09}) are defined as
$\omega^{l,r}_{p}=\omega_0 \pm v_{g} \, p$ where $\omega_0$ is the
central frequency and $v_g > 0$ is the group velocity. This
Hamiltonian is referred to as the ``two-mode" model \cite{fan}.
Throughout the paper we set the Planck's constant $\hbar$ to $1$ and,
thus, measure momentum and energy in wavenumber and frequency units,
correspondingly.

The other constituent of the full Hamiltonian (\ref{eq:eq1}) describes
interaction of the radiation field with the TLS. In the rotating-wave approximation it is given by
\begin{equation} \label{eq:eq3}
  \opH_{int} = g \int \mathd p \left(l^{\dag}_{p} + r^{\dag}_{p}\right) \sigma_{-} + \mathhc,
\end{equation}
where $g$ is the frequency-independent waveguide-qubit coupling strength.

The total number of excitations in the system is defined by the operator
\begin{equation} \label{eq:eq4}
  \hat{N}_{ex} = \int \mathd p \left[l^{\dag}_{p} \, l_{p} + r^{\dag}_{p} \, r_{p}\right] + \sigma_{+} \, \sigma_{-},
\end{equation}
which, similar to (\ref{eq:eq2}), does not contain the interaction terms.
Using the definition (\ref{eq:eq4}) the Hamiltonian ($\ref{eq:eq2}$) is rewritten as
\begin{equation*}
  \opH_0 = \omega_{0} \, \hat{N}_{ex} + \Delta \, \sigma_{+} \, \sigma_{-} + v_{g} \int \mathd p \, p \left[l^{\dag}_{p} \, l_{p} - r^{\dag}_{p} \,
  r_{p}\right],
\end{equation*}
where $\Delta = \omega_{a} - \omega_{0}$ is the detuning between the TLS transition frequency and the central frequency of the waveguide modes.

The operator $\hat{N}_{ex}$ commutes with the Hamiltonian, $\left[\hat{N}_{ex}, \opH \right] \equiv 0$.
Hence, it is the integral of motion. Thus, the system can be equivalently described by the modified Hamiltonian
\begin{equation} \label{eq:eq5}
  \opH' = \opH - \omega_0 \, \hat{N}_{ex} = \opH'_0 + \opH_{int},
\end{equation}
where
\begin{equation*}
  \opH'_0 = v_{g} \int \mathd p \, p \left[l^{\dag}_{p} \, l_{p} -  r^{\dag}_{p} \, r_{p}\right] + \Delta \, \sigma_{+} \, \sigma_{-}.
\end{equation*}
\subsection{The initial state of the radiation field}
We consider the dynamics of light propagating from the left to the right as shown
in Fig. \ref{fig:fig_1}. It is assumed that initially (at $t = t_0$) the qubit is in the ground state and
the propagating light is represented by two single-photon
pulses $|1_{\alpha}\rangle$ and $|1_{\beta}\rangle$. They are
superpositions of single-photon states $l^\dag_p\,\nullket$
weighted by amplitudes $\alpha_p$ and $\beta_p$:
\begin{subequations} \label{eq:eq6}
 \begin{equation} \label{eq:eq6a}
   |1_{\alpha}\rangle = \int \mathd p \, \alpha_p \, l^{\dag}_p(t_0)
   \, \nullket \equiv a^{\dag}_{\alpha} \nullket,
 \end{equation}
 \begin{equation} \label{eq:eq6b}
   |1_{\beta}\rangle = \int \mathd p \, \beta_p \, l^{\dag}_p(t_0) \,
   \nullket \equiv a^{\dag}_{\beta} \nullket,
 \end{equation}
\end{subequations}
where $\nullket$ is the vacuum state of the system and the
factors $\alpha_p$ and $\beta_p$ ensure the normalization
conditions for states (\ref{eq:eq6a}) and (\ref{eq:eq6b})
\begin{equation*}
  \int \mathd p \left|\alpha_p \right|^2 = \int \mathd p \left| \beta_p \right|^2 = 1.
\end{equation*}
It can be verified that $|1_{\alpha,\beta}\rangle$ are the
eigenstates of the operator of the total photon number, $\hat{N}_l =
\int \mathd p \, l^{\dag}_p \, l_p$, with the eigenvalues equal to
unity
\begin{equation*}
  \hat{N}_l|1_{\alpha,\beta}\rangle = 1 \cdot |1_{\alpha,\beta}\rangle,
\end{equation*}
which indicates that $|1_{\alpha,\beta}\rangle$ are the single-photon Fock states.
In what follows we set
\begin{equation} \label{eq:eq7}
  \beta_p = \alpha_p \, \mathe^{\mathi \, p \, L}.
\end{equation}
Then the configuration-space densities of photons in the initial states $|1_{\alpha,\beta}\rangle $ are related as
\begin{equation} \label{eq:eq8}
  \langle 1_{\beta}|\hat{\rho}_l(x,t_0)|1_{\beta}\rangle = \langle 1_{\alpha}|\hat{\rho}_l(x + L, t_0)|1_{\alpha}\rangle,
\end{equation}
where the operator of density of the $l$-photons is given by \cite{sto}
\begin{equation} \label{eq:eq9}
  \hat{\rho}_l(x,t) = \frac{1}{2\pi} \int \mathd p \, \mathd k \, \mathe^{- \mathi \, k \, x} \, l^{\dag}_{p+k/2}(t) \, l_{p-k/2}(t).
\end{equation}
All operators are defined in the Heisenberg representation with the
Hamiltonian given by (\ref{eq:eq5}). It can be seen from
(\ref{eq:eq8}) that the $\beta$- and $\alpha$-pulses being separated
by $L$ have identical shapes. Acting by the raising operators
$a^{\dag}_{\alpha}$ and $a^{\dag}_{\beta}$ on the vacuum state
$\nullket$ we obtain a two-photon Fock state
\begin{equation} \label{eq:eq10}
  |2_{\alpha \beta}\rangle = \nu \, a^{\dag}_{\beta} \, a^{\dag}_{\alpha}\nullket, \quad \nu = \frac{1}{\sqrt{1 + |\chi|^2}}.
\end{equation}
Parameter $\chi = \int \mathd p \, \alpha^{*}_p \, \beta_p$
describes the overlap of the single-photon states (\ref{eq:eq6a}) and (\ref{eq:eq6b}).
When $L = 0$ the constant $\nu$ is equal to $(2!)^{-1/2}$ and
the definition of a two-photon state coincides with the usual
definition of the $n$-photon Fock state given by $|n_{\alpha}\rangle
= (a^{\dag}_{\alpha})^{n}|0\rangle/\sqrt{n!}$ \cite{blow}.

It should be noted that there is another type of two-photon states
named by quantum-correlated photon pairs. They are defined as (see,
for example, \cite{oka})
\begin{equation*}
 |2_{corr}\rangle = \frac{1}{\sqrt{2}} \int \mathd p \int \mathd p' \, \psi(p, p') \, l^{\dag}_{p} \, l^{\dag}_{p'} |0\rangle,
\end{equation*}
where $\psi(p,p') = \psi(p) \, \delta(p + p' - 2 p_0)$ and $2 p_0$
is the total momentum of the photon pair. These states are referred
to as the twin-beam states and can be obtained from spontaneous
parametric down-conversion. The $\delta$-function indicates energy
anticorrelation of two photons.

 We use here the definition (\ref{eq:eq10}). Let us assume that the
ingoing pulses are given by Gaussian distributions. Then $\alpha_p$
is given by
\begin{equation} \label{eq:eq11}
  \alpha_p = \frac{w^{1/2}}{{\pi}^{1/4}} \, \exp\left[-\frac{w^2 \, p^2}{2} - \mathi \, p \, x_{0}\right],
\end{equation}
where $x_0 < 0$. Thus, the single-photon densities are
\begin{equation} \label{eq:eq12}
  \langle 1_{\alpha}|\hat{\rho}_{l}(x, t_0)|1_{\alpha}\rangle = \frac{1}{\pi^{1/2} \, w} \, \mathe^{-(x - x_0)^2/w^2}, \quad
  \langle 1_{\beta}|\hat{\rho}_{l}(x, t_0)|1_{\beta}\rangle = \frac{1}{\pi^{1/2} \, w} \, \mathe^{-(x + L - x_0)^2/w^2}.
\end{equation}
The photon density for the state $|2_{\alpha \beta}\rangle$ is given by
\begin{equation} \label{eq:eq13}
  \langle 2_{\alpha\beta}|\hat{\rho}_{l}(x, t_0)|2_{\alpha\beta}\rangle =
  \frac{\nu^2}{\pi^{1/2} \, w} \, \big[\mathe^{-(x - x_0)^2/w^2} + \mathe^{-(x + L - x_0)^2/w^2} + 2 \mathe^{-L^2/2 \, w^2} \, \mathe^{-(x - x_0 + L/2)^2/w^2}\big],
\end{equation}
where $\nu^{-2} = 1 + \mathe^{-L^2/2 \, w^2}$. For large $L$ the coefficient ${\nu}^{2}$ tends to unity
and the last term in the square brackets which describes the interference effect vanishes.
In this case the incoming field is represented by two independent single-photon pulses.
\begin{figure}[t]
  \centering
  \includegraphics[width=0.5\textwidth]{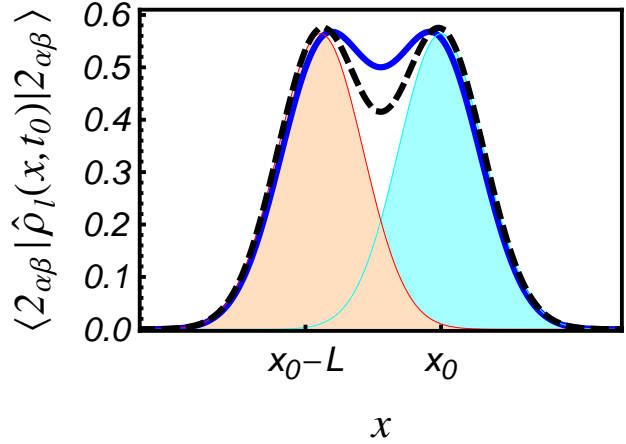}
  \caption{(Color online) The initial photon distributions in the
  configuration space for $L = 2 \, w$. Solid thin lines indicate the densities
  of photons in the states $|1_{\alpha}\rangle$ and $|1_{\beta}\rangle$.
  Dash line is the sum of these densities. Solid thick line is the
  density in the two-photon Fock state $|2_{\alpha\beta}\rangle$.
  The areas under the dash and solid lines are both equal to 2.
  \label{fig:fig_2}}
\end{figure}
It can be seen from Eq. (\ref{eq:eq13}), illustrated  by Fig.
\ref{fig:fig_2}, that the densities at $x = x_0$ and $x = x_0 - L$
are slightly smaller than those given by Eqs.
(\ref{eq:eq12}). On the contrary, the density at the intermediate
position $x = x_0 - L/2$ increases. This variation of the photon
density is caused by interference of the incoming fields. Also, this
phenomenon can be interpreted as a photon-photon interaction caused
by quantum correlations of the incoming pulses.
\section{Equations of motion and evolution of two-photon field} \label{sec:sec_3}
Followed from Hamiltonian (\ref{eq:eq5}) the Heisenberg equations of motion
for photon variables $l_p(t)$ and $r_p(t)$ are as follows
\begin{subequations} \label{eq:eq14}
 \begin{equation} \label{eq:eq14a}
   \left(\partial_{t} + \mathi \, v_{g} \, p \right) l_{p} = - \mathi \, g \, \sigma_{-},
 \end{equation}
 \begin{equation} \label{eq:eq14b}
   \left(\partial_{t} - \mathi \, v_{g} \, p \right) r_{p} = - \mathi \, g \, \sigma_{-},
 \end{equation}
\end{subequations}
with solutions
\begin{subequations} \label{eq:eq15}
 \begin{equation} \label{eq:eq15a}
   l_{p}(t) = \tilde{l}_{p}(t) - \mathi \, g \int^{t}_{t_0} \mathd\tau \, \mathe^{-\mathi \, v_{g} \,  p \, (t - \tau)} \, \sigma_{-}(\tau),
 \end{equation}
 \begin{equation} \label{eq:eq15b}
   r_{p}(t) = \tilde{r}_{p}(t) - \mathi \, g \int^{t}_{t_0} \mathd\tau \, \mathe^{\mathi \, v_{g} \, p \, (t - \tau)} \, \sigma_{-}(\tau),
 \end{equation}
\end{subequations}
where $t > t_0$. Tildes over the operators indicate their free evolution:
\begin{equation} \label{eq:eq16}
  \tilde{l}_{p}(t) = l_{p}(t_0) \, \mathe^{- \mathi \, v_{g} \, p \, (t - t_0)}, \quad
  \tilde{r}_{p}(t) = r_{p}(t_0) \, \mathe^{\mathi \, v_{g} \, p \, (t - t_0)}.
\end{equation}

 Using Hamiltonian (\ref{eq:eq5}), Eqs. (\ref{eq:eq15a}) and (\ref{eq:eq15b}) the equations of motion for the qubit variables take the forms
\begin{equation} \label{eq:eq17}
  \left(\partial_{t} + \Gamma \right) \sigma_{+} \sigma_{-} = \mathi \, g \int
  \mathd p \left(\tilde{l}^{\dag}_{p} + \tilde{r}^{\dag}_{p}\right)\sigma_{-} + \mathhc,
\end{equation}
\begin{equation} \label{eq:eq18}
  \left(\partial_{t} + \mathi \, \Delta + \Gamma/2\right) \sigma_{-} = \mathi \, g
  \left(2 \, \sigma_{+} \sigma_{-} - 1\right) \int \mathd p \left(\tilde{l}_{p} + \tilde{r}_p\right),
\end{equation}
where $\Gamma = 4 \, \pi \, g^2 / v_{g}$. Equation (\ref{eq:eq17}) indicates that parameter $\Gamma$ is a decay rate of the qubit excitation.
Effect of the ingoing field is accounted by the free-moving photon operators $\tilde{l}_{p}$ and $\tilde{r}_{p}$.

\subsection{Phase-space evolution}
The operator of the phase-space distribution function for the $l$-photons is given by \cite{sto}
\begin{equation*}
  \hat{f}_{l}(x, p, t) = \frac{1}{2 \, \pi} \int \mathd k \, \mathe^{-\mathi \, k \, x}
  \, l^{\dag}_{p + k/2}(t) \, l_{p - k/2}(t),
\end{equation*}
which is similar to those used for description of electrons and phonons in semiconductors \cite{chu}.

In order to simplify further considerations we introduce photon
operators $l(x, t)$ and $r(x, t)$ describing, respectively,
annihilation of $l$-photon and $r$-photon at the coordinate $x$.
They are defined as
\begin{equation} \label{eq:eq19}
  l(x,t) = (2 \, \pi)^{-1/2} \int \mathd p \, \mathe^{\mathi \, p \, x} \, l_{p}(t), \quad
  r(x,t) = (2 \, \pi)^{-1/2} \int \mathd p \, \mathe^{\mathi \, p \, x} \, r_{p}(t).
\end{equation}
Substituting expressions (\ref{eq:eq15a}) and (\ref{eq:eq15b}) into Eq. (\ref{eq:eq19}) we obtain the following relations
\begin{subequations} \label{eq:eq20}
 \begin{equation} \label{eq:eq20a}
   l(x, t) = \tilde{l}(x, t) - \sqrt{2 \, \pi} \, \frac{g}{v_g} \, \sigma_{-}(t - \frac{x}{v_g}) \, \theta(x) \, \theta(t - \frac{x}{v_g}),
 \end{equation}
 \begin{equation} \label{eq:eq20b}
   r(x, t) = \tilde{r}(x, t) - \sqrt{2 \, \pi} \, \frac{g}{v_g} \, \sigma_{-}(t + \frac{x}{v_g}) \, \theta(x) \, \theta(t + \frac{x}{v_g}),
 \end{equation}
\end{subequations}
where $\theta(x)$ is the Heaviside step function. As previously, tildes indicate the free-propagating operators which act on the initial state (\ref{eq:eq10}) as
\begin{subequations} \label{eq:eq21}
 \begin{equation} \label{eq:eq21a}
   \tilde{l}(x, t) |2\rangle = (2 \, \pi)^{-1/2}\, \nu \left[A(t - \frac{x}{v})\, |1_{\beta}\rangle + B(t - \frac{x}{v_g}) \, |1_{\alpha}\rangle \right],
 \end{equation}
 \begin{equation} \label{eq:eq21b}
   \tilde{r}(x, t) |2\rangle = 0,
 \end{equation}
\end{subequations}
where $A(t) = \int \mathd p \, \mathe^{-\mathi \, v_g \, p \, t} \, \alpha_p$ and $B(t) = \int \mathd p \, \mathe^{-\mathi \, v_g \, p \, t} \, \beta_p$.
Here and in what follows the index $\alpha\beta$ in the denotation of the initial state $|2_{\alpha\beta}\rangle$ is omitted.

Using (\ref{eq:eq19}) the operator of the phase-space distribution function for the $l$-photons takes the form
\begin{equation} \label{eq:eq22}
 \hat{f}_{l}(x, p, t) = \frac{1}{2 \, \pi} \int \mathd \xi \, \mathe^{\mathi \, p \, \xi} \, l^{\dag}(x + \xi/2,t) \, l(x - \xi/2, t).
\end{equation}
Substituting expression (\ref{eq:eq20a}) into (\ref{eq:eq22}) and
taking into account (\ref{eq:eq21a}) we obtain the average value of
the phase-space distribution for the $l$-photons as
\begin{align} \label{eq:eq23}
  & \langle\hat{f}_{l}(x, p, t)\rangle = \langle\tilde{f}_{l}(x, p, t)\rangle +
  \frac{\Gamma}{2 \, \pi} \int^{2 x /v_g}_{-2 x / v_g} \mathd \tau \, \mathe^{-\mathi
  \, v_g \, p \, \tau} \, \langle \sigma_{+}(t' + \frac{\tau}{2}) \, \sigma_{-}(t' -
  \frac{\tau}{2})\rangle \nonumber \\
  & - \mathi \, \frac{g \, \nu}{2 \, \pi} \int^{2 x/v_g}_{2\left(x/v_g - t\right)}
  \mathd \tau \, \mathe^{\mathi \, v_g \, p \, \tau} \left[A^{*}(t' - \frac{\tau}{2})
  \, \langle 1_{\beta}| + B^{*}(t' - \frac{\tau}{2}) \, \langle 1_{\alpha}|\right]
  \sigma_{-}(t' + \frac{\tau}{2})|2\rangle + \mathcc
  \Big|_{t'=t-x/v_g}.
\end{align}
The first term on the right-hand side of Eq. (\ref{eq:eq23}) describes free propagation of the initial pulse.
The average value $\langle \tilde{f}_l(x,p,t)\rangle$ for distributions (\ref{eq:eq7}) and (\ref{eq:eq11}) is given by
\begin{equation} \label{eq:eq24}
  \langle \tilde{f}_l(x, p, t)\rangle = \frac{\nu^2}{\pi} \, \mathe^{-p^2 \, w^2}
  \left[\mathe^{-X^2(t)/w^2} + \mathe^{-[X(t) + L]^2/w^2} + 2 \, \cos(pL) \, \mathe^{-L^2/4 \, w^2} \, \mathe^{-[X(t) + L/2]^2/w^2}\right],
\end{equation}
where $X(t) = x - x_0 - v_g \, t$. In this case $\langle \tilde{f}_l(x,p,t)\rangle$ depends only on two variables, i.e. $x - v_{g} \, t$ and $p$.
Integration of $\langle \tilde{f}_l(x,p,t)\rangle$ over $p$ gives the configuration-space distribution (\ref{eq:eq13}).
Expression (\ref{eq:eq24}) shows that for the ingoing Gaussian pulse the initial phase-space distribution is positive at any point of phase space.

\begin{figure}[t]
  \centering
  \includegraphics{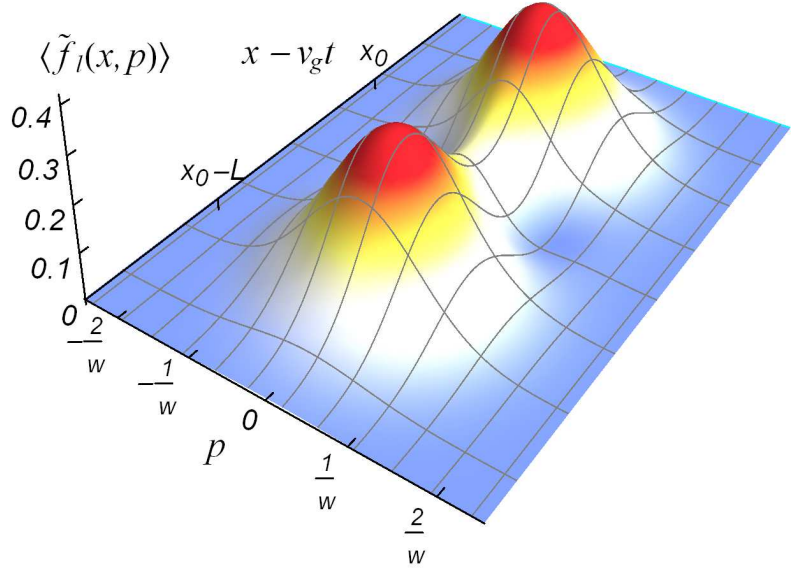}
  \caption{(Color online) The initial photon phase-space distribution for $L = 3 w$ exhibiting two-peak structure.
  Oscillations in the vicinity of $X(t) = - L/2$ are damped.
  \label{fig:fig_3}}
\end{figure}

Figure \ref{fig:fig_3} displays the initial phase-space distribution
for $L = 3 w$. This distribution exhibits a two-peak structure with
maxima at $x_0$ and $x_0 - L$ as it follows from (\ref{eq:eq24}).
The interference of the single-photon wavepackets is described by
the third term in the brackets in Eq. (\ref{eq:eq24}). For larger
$L$ the interference becomes less pronounced and the initial
distribution tends to form two solitary peaks. For $L = 0$
the inital distribution has the only maximum at $x_0$.

The second and third terms on the right-hand side of Eq. (\ref{eq:eq23})
arise due to interaction of the ingoing pulse with the qubit. These
terms are nonzero only for $x > 0$. The integration limits are
imposed by the $\theta$-functions in Eqs. (\ref{eq:eq20a}) and
(\ref{eq:eq20b}). The second term in (\ref{eq:eq23}) describes the
$l$-mode of the field re-emitted by the TLS. The third term on the
right-hand side of (\ref{eq:eq23}) is linear with respect to the
waveguide-qubit coupling parameter $g$. This term describes
interference of the ingoing field and the field re-emitted by the
TLS.

The operator of the phase-space distribution for the $r$-photons, $\hat{f}_{r}(x, p, t)$,
is defined by replacing $l$ with $r$ in (\ref{eq:eq22}).
Taking into account Eqs. (\ref{eq:eq20b}) and (\ref{eq:eq21b}) we obtain
$\langle \hat{f}_{r}(x, p, t)\rangle$ as
\begin{equation} \label{eq:eq25}
 \langle \hat{f}_r(x, p, t)\rangle = \frac{\Gamma}{2 \, \pi}
 \int^{-2 x /v_g}_{2 x / v_g} \mathd \tau \, \mathe^{\mathi \,
  v_g \, p \, \tau} \, \langle \sigma_{+}(t' + \frac{\tau}{2}) \,
  \sigma_{-}(t' - \frac{\tau}{2})\rangle\Big|_{t' = t + x/v_g}.
\end{equation}
This distribution is nonzero for $x < 0$ and coincides with the
third term on the right-hand side of Eq. (\ref{eq:eq23}), with $v_g$
is replaced by $-v_g$, due to the symmetry properties of the considered
system. Expression (\ref{eq:eq25}) shows that the reflected field
consists only of the $r$-mode of the field re-emitted by the TLS.

As seen in Eqs. (\ref{eq:eq23}) and (\ref{eq:eq25}), in order to
calculate the photon phase-space distributions we should know
two-time correlator $\langle\sigma_{+}(t) \, \sigma_{-}(t')\rangle$
and matrix elements $\langle 1_{\alpha, \beta}|\sigma_{-}(t)|2\rangle$.
As follows from (\ref{eq:eq18}) evolution of $\langle \sigma_{+}(t)
\, \sigma_{-}(t')\rangle$ is governed by the equation
\begin{align} \label{eq:eq26}
 & \left(\partial_t - \mathi \, \Delta + \Gamma/2 \right) \langle
 \sigma_{+}(t) \, \sigma_{-}(t') \rangle = \mathi \, g \, \nu \,
 \Big[A^{*}(t)\langle 1_{\beta}| + B^{*}(t)\langle 1_{\alpha}| \nonumber \\
 & - 2 \, A^{*}(t)\langle 1_{\beta}|\sigma_{+}(t)|0\rangle
 \langle 0|\sigma_{-}(t) - 2 \, B^{*}(t)\langle 1_{\alpha}|
 \sigma_{+}(t)|0\rangle\langle 0|\sigma_{-}(t)\Big]
 \sigma_{-}(t')|2\rangle .
\end{align}
Equation of motion for matrix element $\langle 0|\sigma_{-}(t) \,
 \sigma_{-}(t')|2\rangle$ in the right-hand side of Eq. (\ref{eq:eq26}) is given by
\begin{equation} \label{eq:eq27}
 \left(\partial_t + \mathi \, \Delta + \Gamma/2\right) \langle 0|\sigma_{-}(t) \,
 \sigma_{-}(t')|2\rangle = - \mathi \, g \, \nu \left[A(t) \, \langle 0
 |\sigma_{-}(t')|1_{\beta}\rangle + B(t) \, \langle
 0|\sigma_{-}(t')|1_{\alpha}\rangle\right].
\end{equation}
The property $\sigma_{-}(t)|1_{\alpha, \beta}\rangle=\left[\langle
1_{\alpha, \beta}|\sigma_{+}(t)\right]^{\dag}$ and the relation
\begin{equation} \label{eq:eq28}
 \sigma_{-}(t)|1_{\alpha, \beta}\rangle = \langle 0|\sigma_{-}(t)
 |1_{\alpha, \beta}\rangle |0 \rangle, \quad\langle 1_{\alpha, \beta}
 |\sigma_{+}(t) = \langle 0|\sigma_{-}(t)|1_{\alpha, \beta}\rangle^\ast \langle 0|
\end{equation}
derived in Appendix \ref{sec:app_a} are utilized to obtain the right-hand side of Eqs.
(\ref{eq:eq26}) and (\ref{eq:eq27}).
The explicit expression for $\langle 0|\sigma_{-}|1_{\alpha, \beta}\rangle$
follows from Eqs. (\ref{eq:eq18}) and (\ref{eq:pr1}) and has the form
\begin{equation} \label{eq:eq29}
 \langle 0|\sigma_{-}(t)|1_{\alpha}\rangle = - \mathi \, g \int^{t}_{0} \mathd \tau \,
 \mathe^{-(\mathi \, \Delta + \Gamma/2)(t - \tau)} \, A(\tau).
\end{equation}
The expression for $\langle 0|\sigma_{-}(t)|1_{\beta}\rangle$ is obtained
by replacing $A(t)$ with $B(t)$. For the sake of brevity, hereinafter we set $t_0 = 0$.

It is assumed that $t > t'$ in Eq. (\ref{eq:eq26}). (For $t < t'$ the
relation $\langle \sigma_{+}(t') \, \sigma_{-}\rangle = \langle \sigma_{+}(t)
  \, \sigma_{-}(t')\rangle^{*}$ is used.)
Thus, for $t = t'$ we get the inital conditions $\langle \sigma_{+}(t) \,
\sigma_{-}(t')\rangle|_{t=t'} = \langle \sigma_{+} \, \sigma_{-} \rangle|_{t}$
and $\langle 0|\sigma_{-}(t) \, \sigma_{-}(t')|2\rangle|_{t=t'} = 0$.
Taking into account Eqs. (\ref{eq:eq17}) and (\ref{eq:eq18}) the equations of
motion for $\langle \sigma_{+} \, \sigma_{-} \rangle$ and $\langle 1_{\alpha,
\beta}|\sigma_{-}|2\rangle$ form a full set of equations
\begin{subequations} \label{eq:eq30}
\begin{equation} \label{eq:eq30a}
 \left(\partial_t + \Gamma\right)\langle \sigma_{+} \, \sigma_{-} \rangle = i
 \, g \left[A^{*}(t)\langle 1_{\beta}|\sigma_{-}|2\rangle + B^{*}(t)\langle
 1_{\alpha}|\sigma_{-}|2\rangle \right] + \mathcc,
\end{equation}
\begin{align} \label{eq:eq30b}
 \left(\partial_t + \mathi \, \Delta + \Gamma/2\right) \langle 1_{\alpha}|\sigma_{-}(t)
 |2\rangle = \, & 2 \, \mathi \, g \, \nu \, \langle 1_{\alpha}|\sigma_{+}(t)|0\rangle
  \left[A(t) \, \langle 0|\sigma_{-}(t)|1_{\beta}\rangle + B(t) \, \langle 0
  |\sigma_{-}(t)|1_{\alpha}\rangle\right] \nonumber \\
 & - \mathi \, g \, \nu \left[\chi \, A(t) + B(t)\right].
\end{align}
\end{subequations}
Equation for $\langle 1_{\beta}|\sigma_{-}(t)|2\rangle$ can be
obtained by mutual replacement of $\alpha$ and $\beta$ as well as $A(t)$ and $B(t)$
in Eq. (\ref{eq:eq30b}). Generalization of the described
scheme for the case of an $n$-photon Fock state is presented in
Appendix \ref{sec:app_b}.

Figure \ref{fig:fig_4} shows phase-space distributions of
photons after their interaction with the TLS for different values of
$L$ and $\Gamma$. In contrast to the positive initial
distribution (\ref{eq:eq24}),  the phase-space distribution of
transmitted photons ($x>0$) exhibits a distinct ``dip" which
can form an area of negative values. This is the result of
anticorrelation between  the ingoing and re-emitted fields.
When single-photon components of the ingoing state (\ref{eq:eq10})
have significant overlap, the ``dip" in the phase-space distribution
is less pronounced than in the case of
single-photon input considered in Ref. \cite{sto}. With increase of
$\Gamma$ the negative regions in the phase-space distribution
vanish. The reason for it is that for greater coupling $g$
the qubit is excited more effectively and the term describing the
TLS re-emission in Eq. (\ref{eq:eq23}) dominates the interference
term. With increase of the spatial separation $L$ the
interference between the initial pulses  decays. For large $L$ the
problem reduces to the scattering of independent single-photon pulses.
\begin{figure}[t]
  \centering
  \includegraphics[width = 0.9\textwidth]{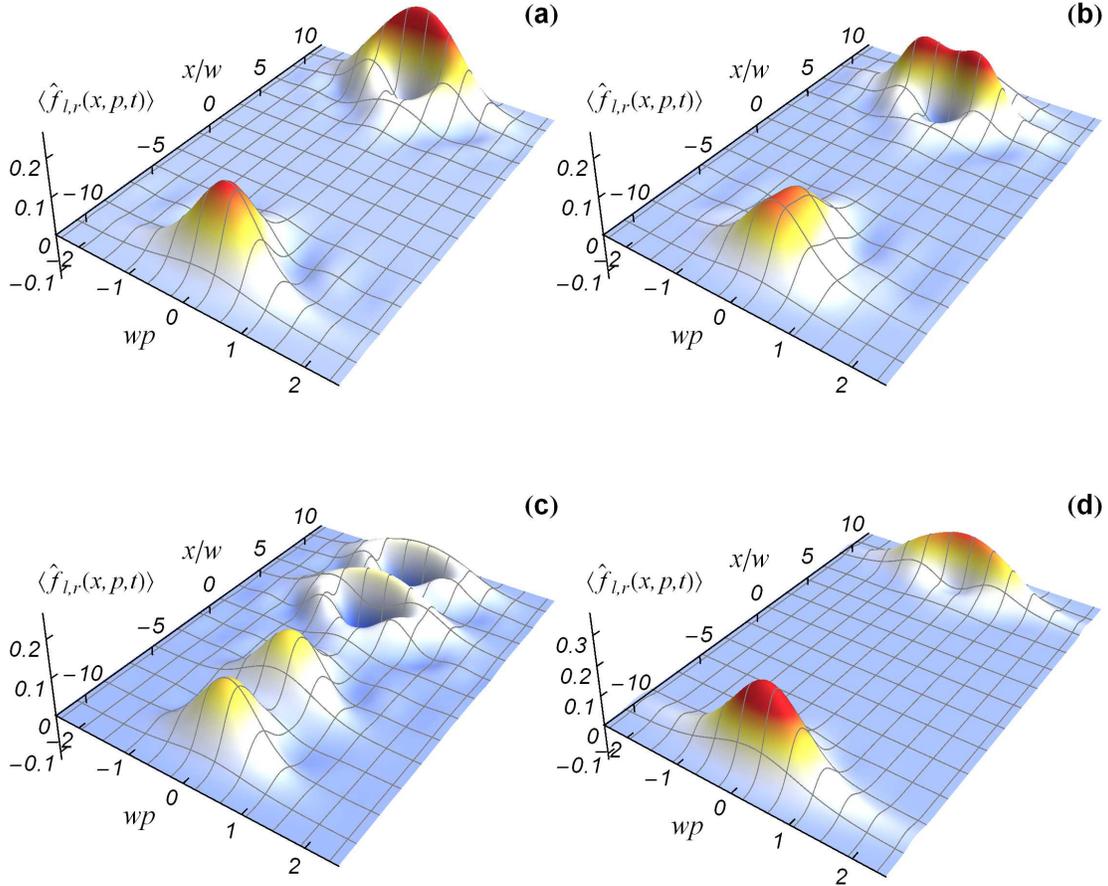}
  \caption{(Color online) Phase-space distribution for photons after interaction
  with TLS for different $\Gamma$ and $L$.
  All calculations are performed for $t = (10 \, w + |x_0|)/v_g$, $x_0 = -10
  \, w$ and $\Delta = 0$.
  The other parameters are the following: (a) $\Gamma = \Omega$, $L = 0$;
  (b) $\Gamma = \Omega$, $L = 2w$; (c) $\Gamma = \Omega$, $L = 5 w$; (d) $\Gamma = 2 \Omega$, $L = 0$.
  Phase-space distributions (a)-(c) exhibit negative values while distribution (d) does not.
  Parameter $\Omega = v_{g}/w$ is the bandwidth of the ingoing pulses.
 \label{fig:fig_4}}
 \end{figure}

\subsection{Photon densities and spectra}
\begin{figure}[t]
  \centering
  \includegraphics{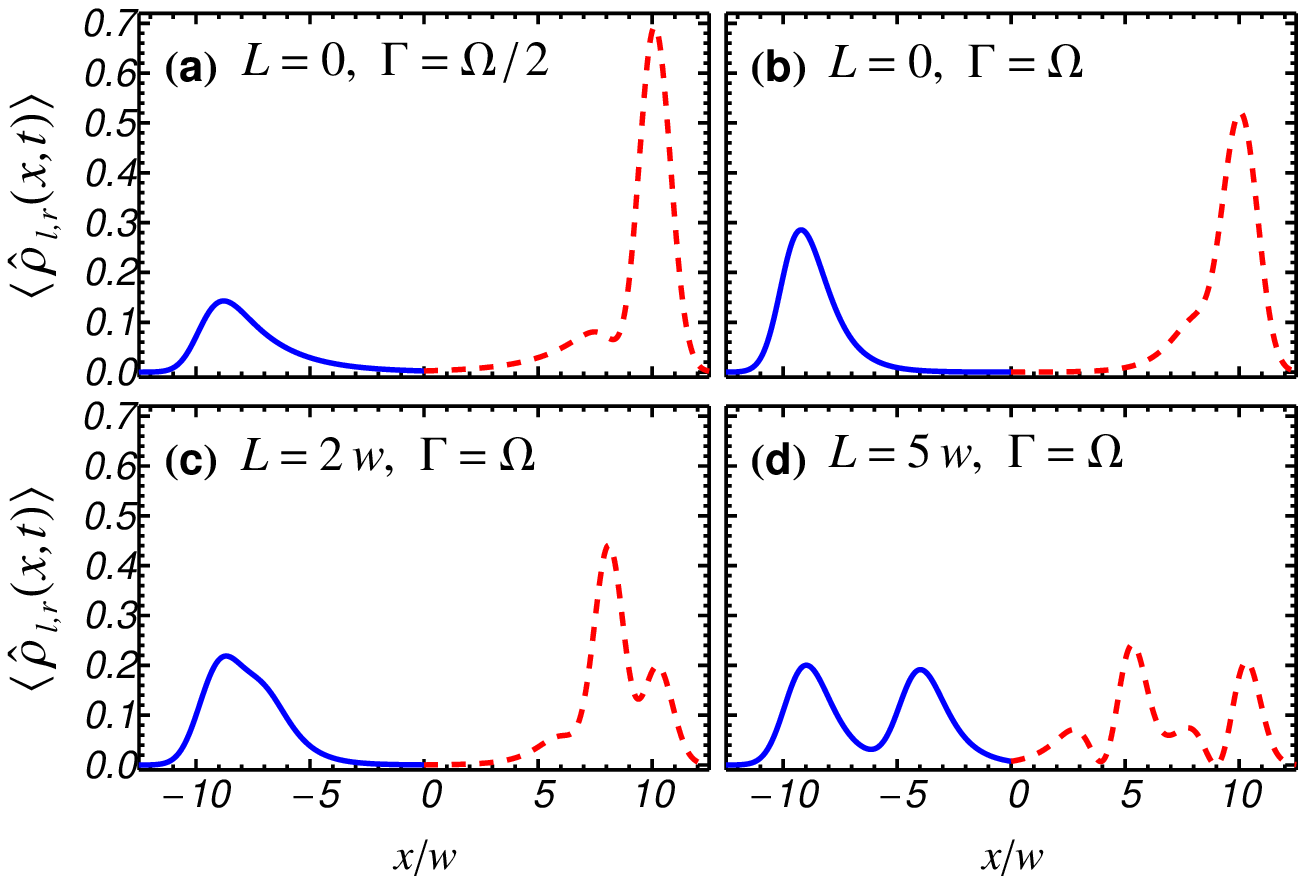}
  \caption{(Color online) Configuration-space densities of the transmitted
  (dash red lines) and reflected (solid blue lines) photons.
  The rest of the parameters are the same as in Fig. \ref{fig:fig_4}. \label{fig:fig_5}}
\end{figure}

The average photon densities $\langle \hat{\rho}_{l,r}(x,t)\rangle$ can be obtained
by integrating the phase-space distribution functions over all momenta
$\langle \hat{\rho}_{l, r}(x, t)\rangle = \int \mathd p \, \langle \hat{f}_{l, r}(x, p, t)\rangle$.
The results of calculation of $\langle \hat{\rho}_{l,r}(x,t)\rangle$ are shown in Fig \ref{fig:fig_5}.

Increase of the reflection can be seen if the qubit-waveguide
coupling $\Gamma$ or pulse width $w$ increases (see, for example,
Refs. \cite{dom, zheng, sto}). Stronger waveguide-qubit
coupling (or longer ingoing pulses) results in greater probability
of the TLS to be excited that leads to more pronounced
destructive interference effects in the transmitted field.

\begin{figure}[t]
  \centering
  \includegraphics{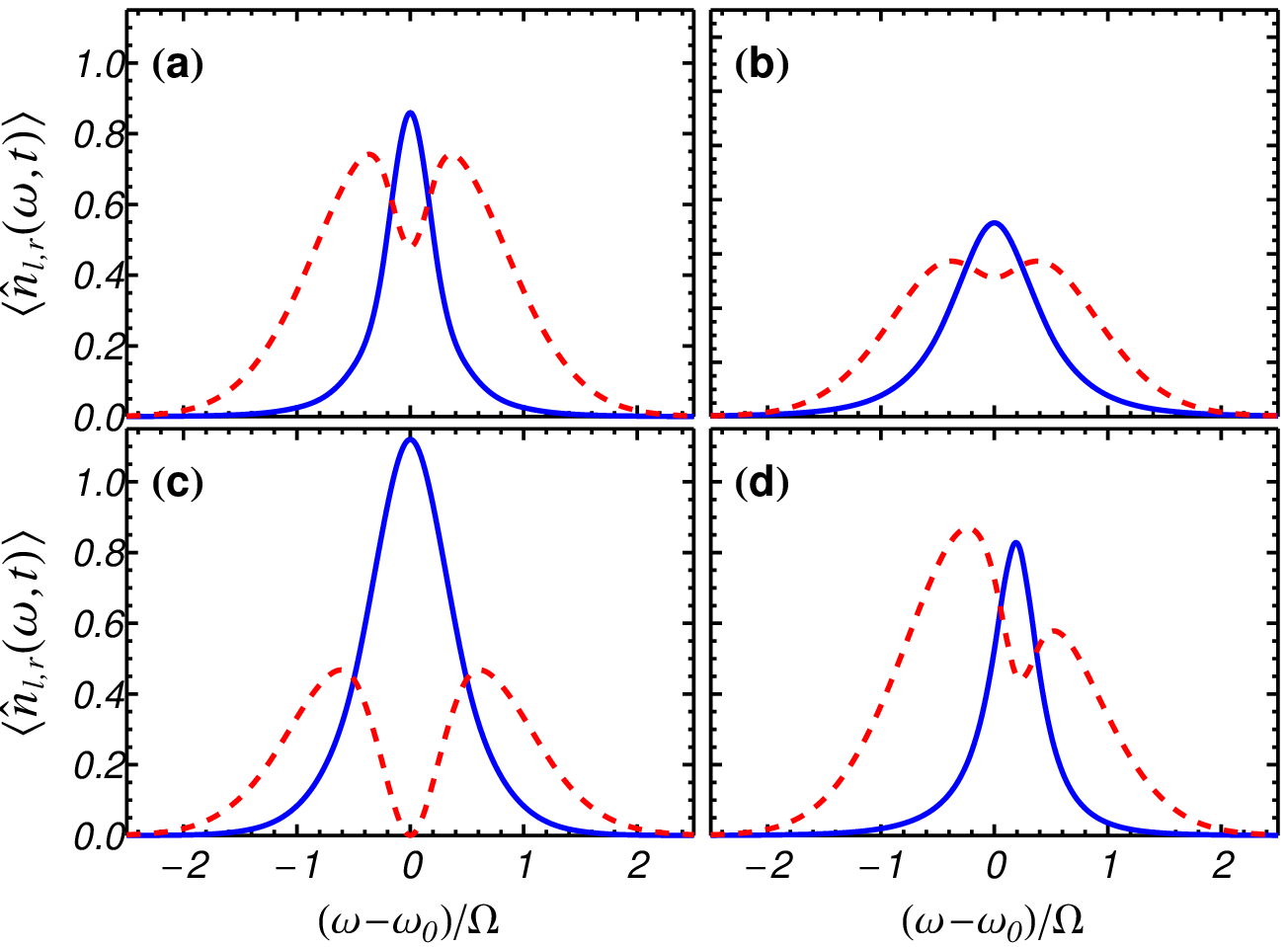}
  \caption{(Color online) Spectra of the transmitted (red dash lines) and
  reflected (blue solid lines) light. Parameters are the following:
  (a) $\Gamma = \Omega/2$, $L = 0$, $\Delta = 0$;
  (b) $\Gamma = \Omega$, $L = 0$, $\Delta = 0$;
  (c) $\Gamma = \Omega$, $L = 10 w$, $\Delta = 0$;
  (d) $\Gamma = \Omega/2$, $L = 0$, $\Delta = \Omega/5$.
  The calculations are performed for $t \gg \Gamma^{-1} + |x_0|/v_g$
  that ensures the outgoing pulses to be far from the TLS; $x_0 = -10 \, w$. \label{fig:fig_6}}
 \end{figure}

Similar reasonings, but expressed in different terms, are applicable
for explanation of the features of the outgoing photon spectra.
Integration of the phase-space distribution over spatial
variable gives the momentum-space distribution. For linear
dependencies of $\omega^{l,r}$ on $p$ the relations between the
photon momenta and frequency are given by $p = \pm (\omega -
\omega_0)/v_g$, where the sign``$+$"(``$-$"") corresponds to the $l$-mode
($r$-mode), respectively.
Thus, the spectra of the outgoing light are determined as
\begin{equation*}
 \langle \hat{n}_{l,r}(\omega, t)\rangle =
 \int \mathd x \, \langle \hat{f}_{l,r}(x, \pm (\omega - \omega_0)/v_g , t)\rangle.
\end{equation*}
Figure \ref{fig:fig_6} represents the outgoing light spectra
for different parameters of the system.

The TLS emission spectrum has a maximum at $\omega_a$ with linewidth
$\Gamma$. Thus, the maximal TLS excitation and reflection occurs at
resonance $\omega_0 = \omega_a$. If the bandwidth of the ingoing
wavepacket is larger than the linewidth of the TLS emission, only
the frequencies close to the resonance $\omega - \omega_0 = \Delta$
provide an effective photon-TLS interaction. The portions of the
ingoing wavepacket with frequencies far from the resonance pass the
TLS almost freely. That is why the spectrum of the reflected light
has width $\Gamma$ and maximum at $\omega - \omega_0 = \Delta$. The
transmitted light spectrum has a pronounced minimum at this point.
The TLS operates here as a quantum spectral filter resembling a
band-stop filter in radioelectronics. For $\Delta = 0$ the spectra
of the reflected and transmitted light are symmetric with respect to
the point $\omega - \omega_0 = 0$. For $\Delta \neq 0$ the spectra
become asymmetric. When the ingoing state consists of two
strongly-overlapping components, $\langle
\hat{n}_{l}(\omega)\rangle$ does not drop to zero at
$\omega-\omega_0=\Delta$ while for the single-photon input
$\langle\hat{n}_{l}(\omega)\rangle = 0$ at this frequency (see Ref.
\cite{sto}). This is because only one photon can be absorbed by the
TLS at a moment.
\section{Photon statistics} \label{sec:sec_4}
Photon number fluctuations of the outgoing light are described by
the variances $\langle \delta\hat{N}^2_{l,r}\rangle =
\langle(\hat{N}_{l,r} - \langle\hat{N}_{l,r}\rangle)^2\rangle =
\langle\hat{N}_{l, r}^2\rangle - \langle\hat{N}_{l,r}\rangle^2$. The
further consideration  is for $t \gg |x_0|/v_g + \Gamma^{-1}$ when
the outgoing pulses are far from the TLS. In this case the average
numbers of reflected and transmitted photons do not depend on time.
They are connected by the relation
\begin{equation} \label{eq:eq31}
 \langle\hat{N}_{l}\rangle = \langle\hat{N}_{0}\rangle - \langle\hat{N}_{r}\rangle,
\end{equation}
where $\hat{N}_0 = \hat{N}_{l}(t=0)$ is the operator of the number
of ingoing photons. Using relation (\ref{eq:eq31}) and taking into
account $\langle\delta N_0^2\rangle = 0$ for any Fock state we
obtain that variances of the reflected and transmitted photon
numbers are equal: $\langle \delta \hat{N}_{l}^2 \rangle = \langle
\delta \hat{N}_{r}^2\rangle$. To calculate $\langle
\delta\hat{N}^2_{r}\rangle$ the values of
$\langle\hat{N}^2_{r}\rangle$ and $\langle\hat{N}_{r}\rangle$ are
required. The average number of the  reflected photons is defined by
\begin{equation*}
 \langle\hat{N}_{r}\rangle = \int \mathd x \int \mathd p \, \langle\hat{f}_{r}(x,
 p)\rangle,
\end{equation*}
which with the help of Eq. (\ref{eq:eq25}) gives
\begin{equation} \label{eq:eq32}
 \langle\hat{N}_{r}\rangle = \frac{\Gamma}{2}\int^{t}_{0} \mathd \tau \, \langle \sigma_{+}
 \, \sigma_{-}\rangle_{\tau}.
\end{equation}
The integrand in (\ref{eq:eq32}) is governed by Eq.
(\ref{eq:eq30a}). For $\hat{N}^2_{r}$ we can use the representation
$\hat{N}^2_{r} = \int \mathd x_1 \int \mathd x_2 \, r^{\dag}(x_1) \,
r^{\dag}(x_2) \, r(x_2) \, r(x_1) + \hat{N}_{r}$. Taking into
account Eqs. (\ref{eq:eq20b}) and (\ref{eq:eq21b}) we obtain
\begin{equation*}
 \langle\hat{N}^2_{r}\rangle = \frac{\Gamma^2}{4} \int^{t}_{0} \mathd \tau_1 \int^{t}_{0} \mathd \tau_2 \, \langle\sigma_{+}(\tau_1) \, \sigma_{+}(\tau_2) \, \sigma_{-}(\tau_2) \, \sigma_{-}(\tau_1)\rangle + \langle\hat{N}_{r}\rangle.
\end{equation*}
Using the property
\begin{equation} \label{eq:eq33}
 \sigma_{-}(t) \, \sigma_{-}(t')|2\rangle = \langle 0|\sigma_{-}(t) \,
 \sigma_{-}(t')|2\rangle|0\rangle ,
\end{equation}
derived in Appendix \ref{sec:app_a}, we get
\begin{equation} \label{eq:eq34}
 \langle\hat{N}_{r}^2\rangle = \frac{\Gamma^2}{4} \int^{t}_{0}
 \mathd \tau \int^{t}_{0} \mathd \tau' \, \left|\langle 0|\sigma_{-}(\tau)
 \, \sigma_{-}(\tau')|2\rangle\right|^2 + \langle \hat{N}_{r} \rangle.
\end{equation}
The matrix element $\langle 0|\sigma_{-}(t) \,
\sigma_{-}(t')|2\rangle$ in (\ref{eq:eq34}) obeys Eq. (\ref{eq:eq27}).

\begin{figure}[t]
  \centering
  \includegraphics[width=0.9\textwidth]{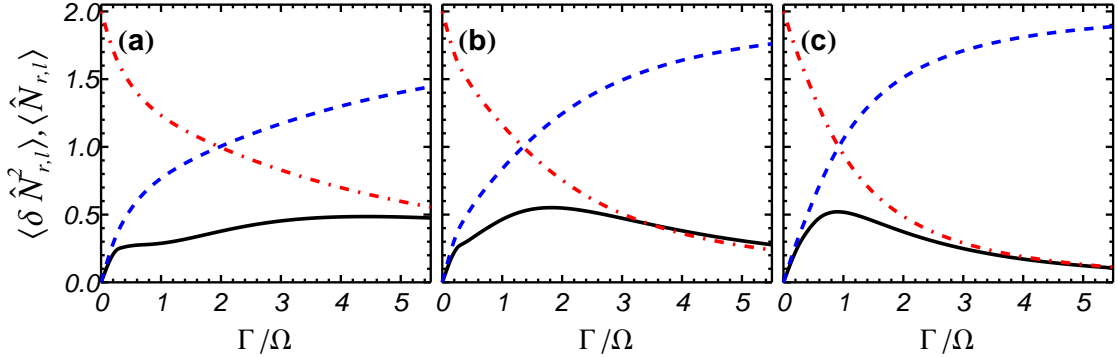}
  \caption{(Color online) Variances of outgoing photon numbers
   (solid black lines) and the average numbers of reflected (dash blue lines)
   and transmitted (dot-dash red lines) photons
   vs. $\Gamma$ at $\Delta = 0$ for different $L$:
   $L = 0$ (a), $L = 2 w$ (b) and $L = 5 w$ (c).
   \label{fig:fig_7}}
\end{figure}

The results of calculations shown in Fig. \ref{fig:fig_7} confirm
the tendency of reflectance to increase when the
waveguide-TLS coupling increases. This tendency becomes more
pronounced for greater $L$. Also we can see that
the variance of the reflected photons is less than the
average photon number for any $\Gamma$ and $L$.
This manifests the sub-Poissonian statistics (antibunching)
of the reflected photons which are emitted one by one by the TLS.
In contrast, the transmitted light can exhibit super-Poissonian statistics
(bunching).
\section{Summary} \label{sec:summ}
Our approach, based on the formalism of the phase-space distribution
function, provides a detailed picture of interaction of two-photon
pulses with TLS. It makes it possible to describe not only the
asymptotic characteristics of the outgoing light, such as
transmission/reflection coefficients or photon scattering
probabilities \cite{sfan, xufan, zheng}, but also to investigate the
dynamics of the whole system (see Fig. \ref{fig:fig_4}). A full set
of equations describing the evolution of the two-photon state is
derived and solved for different parameters. It is shown that along
with the coupling strength $g$ and the initial pulse width the
spatial separation between the single-photon components of the
ingoing field strongly affects the dynamics of the system.

The method of photon phase-space operator has an advantage of high
universality. Its integration over the momentum $p$ results in the
photon density in the configuration space (see Fig. \ref{fig:fig_5}).
Similarly, integration of the phase-space
distribution over the configuration space gives light spectra.
Our calculations show that spectra of the reflected and transmitted
photons have distinct difference due to peculiarities of
the TLS response. Owing to the saturable behavior of the TLS excitation
the spectra of the outgoing light for the two-photon input
differ from those for the single-photon input (see Ref. \cite{sto}).

We have studied photon number fluctuations of the outgoing
light. The corresponding variances for both modes are found to be equal for any
$n$-photon Fock state. These variances determine signal/noise ratios
that describe the possibility of outgoing light to be utilized.
Dependence of the variances on the coupling parameter $g$ and the
separation distance $L$ is analyzed. Our calculations reveal the
antibunching statistics of the reflected photons regardless of the
choice of $\Gamma$ and $L$. By contrast, the statistics of
transmitted photons can be sub-Poissonian or super-Poissonian depending on
choice of the parameters $\Gamma$ and $L$.

To summarize, tuning both the waveguide-TLS coupling and the ingoing
pulse separation can control spatio-temporal and statistical
characteristics of the outgoing light that may find applications in
QIP.
\begin{acknowledgments}
 The authors thank V. Bondarenko, S. Lukyanets and A. Semenov for useful discussions and comments.
\end{acknowledgments}
\appendix
\section{Operator properties} \label{sec:app_a}
\subsection{Properties of free-moving operators}
Here we prove the commutation relations
\begin{equation} \label{eq:comm1}
 \left[\int \mathd p \, \tilde{l}_p(t), \sigma_{-}(t')\right] = \left[\int \mathd p \, \tilde{r}_p(t),
 \sigma_{-}(t')\right] = 0, \quad t \geq t'.
\end{equation}
Using the representation $\tilde{l}_{p}(t) = \tilde{l}_{p}(t') \, \mathe^{-\mathi \, v_g \, p \, (t - t')}$,
equal-time commutator $[l_p, \sigma_{-}] = 0$ and Eq. (\ref{eq:eq15a}) we obtain
\begin{align*}
 \left[\int \mathd p \, \tilde{l}_p(t), \sigma_{-}(t')\right] & = \int \mathd p \, \mathe^{-\mathi \, v_g \, p
 \, (t - t')} \left[l_{p}(t') + \mathi \, g \int^{t'}_{0} \mathd \tau\, \mathe^{-\mathi \, v_g \, p
 \, (t' - \tau)} \, \sigma_{-}(\tau) , \sigma_{-}(t')\right] \\
 & = \mathi \, g \int \mathd p \int^{t'}_{0} \mathd \tau \, \mathe^{-\mathi \, v_g \, p \, (t -
 \tau)} \left[\sigma_{-}(\tau), \sigma_{-}(t') \right] = \mathi \, \frac{2\pi \, g}{v_g}
 \left[\sigma_{-}(t), \sigma_{-}(t')\right] \theta(t' - t).
\end{align*}
Presence of the $\theta$-function in the last term shows that the
initial expression is equal to zero if $t>t'$. When
$t=t'$ the commutator $\left[\sigma_{-}(t),
\sigma_{-}(t')\right]$ is equal to zero.
This proves Eq. (\ref{eq:comm1}).
Similar reasonings are applicable for the second commutator in
(\ref{eq:comm1}).

It follows directly from the definitions of the free-moving operators
(\ref{eq:eq15}), the single-photon states (\ref{eq:eq6}) and the two-photon states (\ref{eq:eq10}) that
\begin{equation} \label{eq:pr1}
  \int \mathd p \, \tilde{l}_{p}(t) |1_{\alpha}\rangle = A(t)|0\rangle, \quad \int \mathd p \, \tilde{l}_{p}(t)|1_{\beta}\rangle = B(t)|0\rangle,
\end{equation}
\begin{equation} \label{eq:pr2}
   \int \mathd p \, \tilde{l}_{p}(t) |2\rangle = \nu \left[A(t)|1_{\beta}\rangle + B(t)|1_{\alpha}\rangle\right]
\end{equation}
and
\begin{equation} \label{eq:pr3}
 \tilde{r}_{p}(t)|2\rangle = \tilde{r}_{p}(t) |1_{\alpha}\rangle = \tilde{r}_{p}(t) |1_{\beta}\rangle = 0.
\end{equation}
These relations are widely used in the paper.
\subsection{Derivation of the relations (\ref{eq:eq28}) and ({\ref{eq:eq33}})}
The action of $\sigma_{-}(t)$ on the states $|1_{\alpha,\beta}\rangle$ gives the state $C_{\alpha, \beta}(t)|0\rangle$.
To prove this we use the solution of Eq. (\ref{eq:eq18}):
\begin{equation} \label{eq:A1}
 \sigma_{-}(t) = \tilde{\sigma}_{-}(t) + \mathi \, g \int_{t_0}^t \mathd \tau \,
 \mathe^{-(\mathi \, \Delta + \Gamma/2)(t - \tau)} \, (2 \, \sigma_{+} \, \sigma_{-}|_{\tau} - 1) \int
 \mathd p \, (\tilde{l}_p + \tilde{r}_p)|_{\tau}.
\end{equation}
With account for the relation $\sigma_{-}(t)|0\rangle = \tilde{\sigma}_{-}(t)|1_{\alpha,\beta}\rangle = 0$ and Eq. (\ref{eq:A1}), we have
\begin{equation} \label{eq:A2}
 \sigma_{-}(t)|1_{\alpha}\rangle = - \mathi \, g \int_{t_0}^t \mathd \tau \, \mathe^{-(\mathi \, \Delta + \Gamma/2)(t - \tau)} \, A(\tau)|0\rangle.
\end{equation}
The expression for the state  $|1_\beta\rangle$ can be obtained from (\ref{eq:A2}) by replacing $A(t)$ with $B(t)$.

As we see, the action of the lowering operator $\sigma_{-}(t)$
on the single-photon state $|1_{\alpha,\beta}\rangle$ moves the system into the vacuum state $|0\rangle$
\begin{equation}\label{eq:A3}
 \sigma_{-}(t) |1_{\alpha,\beta}\rangle = C_{\alpha, \beta}(t) |0\rangle,
\end{equation}
where the factor $C_{\alpha, \beta}(t)$ is given by (\ref{eq:A2}).
It follows from Eq. (\ref{eq:A3}) that the state $\sigma_-(t)|1_{\alpha,\beta}\rangle$ can be represented in the equivalent form
\begin{equation}\label{eq:A4}
 \sigma_{-}(t)|1_{\alpha,\beta}\rangle = \langle 0|\sigma_{-}(t)|1_{\alpha,\beta}\rangle |0\rangle,
\end{equation}
where $\langle 0|\sigma_-(t)|1_{\alpha,\beta}\rangle \equiv C_{\alpha, \beta}(t)$.

Similarly we can prove that $\sigma_{-}(t) \, \sigma_{-}(t')|2\rangle = \langle 0|\sigma_{-}(t) \, \sigma_{-}(t')|2\rangle|0\rangle$.
Taking into account (\ref{eq:comm1}), (\ref{eq:pr2}) and (\ref{eq:pr3}) the equation of motion for $\sigma_{-}(t) \, \sigma_{-}(t')|2\rangle$ for $t > t'$ has the form
\begin{equation}
 \left(\partial_{t} + \mathi \, \Delta + \Gamma/2\right)\sigma_{-}(t) \, \sigma_{-}(t')|2\rangle = \mathi \, g \, \nu \left(2 \, \sigma_{+} \, \sigma_{-}|_{t} - 1\right) \sigma_{-}(t')\left[A(t)|1_{\beta}\rangle + B(t)|1_{\alpha}\rangle\right],
\end{equation}
with solution
\begin{equation*}
 \sigma_{-}(t) \, \sigma_{-}(t')|2\rangle = \mathi \, g \, \nu \left(2 \, \sigma_{+} \, \sigma_{-}|_{t} - 1\right) \int^{t}_{t'} \mathd \tau \, \mathe^{(-\mathi \, \Delta + \Gamma/2)(t-\tau)} \, \sigma_{-}(\tau)\left[A(\tau)|1_{\beta}\rangle + B(\tau)|1_{\alpha}\rangle\right].
\end{equation*}
Due to relation (\ref{eq:A4}) and property $\sigma_{+} \, \sigma_{-}|0\rangle = 0$ we obtain
\begin{equation}
 \sigma_{-}(t) \, \sigma_{-}(t')|2\rangle = - \mathi \, g \, \nu \int^{t}_{t'} \mathd \tau \, \mathe^{(-\mathi \, \Delta + \Gamma/2)(t-\tau)} \left[A(\tau) \, C_{\beta}(\tau) + B(\tau) \, C_{\alpha}(\tau)\right]|0\rangle.
\end{equation}
Thus, the state $\sigma_{-}(t) \, \sigma_{-}(t')|2\rangle$ can be represented as
\begin{equation}
 \sigma_{-}(t) \, \sigma_{-}(t')|2\rangle = \langle 0|\sigma_{-}(t) \, \sigma_{-}(t')|2\rangle|0\rangle.
\end{equation}
\section{The {n}-photon Fock state} \label{sec:app_b}
The previous considerations can be extended for the $n$-photon Fock
state. Its general form is given by
\begin{equation} \label{eq:b1}
 |n_{\{\alpha_{j}\}}\rangle = \nu_{n} \prod_{j=1,n} a^{\dag}_{\alpha_{j}}|0\rangle,
\end{equation}
where $\nu_{n}$ is the normalization constant.
For simplicity, we consider $\alpha_{j} = \alpha$ and the state (\ref{eq:b1}) reduces to
\begin{equation} \label{eq:b2}
  |n_{\alpha}\rangle = \frac{1}{\sqrt{n!}} \left[a^{\dag}_{\alpha}\right]^{n} \nullket.
\end{equation}
For the state (\ref{eq:b2}) the initial configuration-space density
of photons is given by the single-peak distribution
\begin{equation*}
 \langle n|\hat{\rho}(x, t = 0)|n \rangle = \frac{n}{\sqrt{\pi} \, w} \, \mathe^{-(x - x_0)^2/w^2}
\end{equation*}
that coincides with the density in the two-photon state
(\ref{eq:eq10}) when $n = 2$, $L = 0$, and $|1_\alpha\rangle =
|1_\beta\rangle$.

Let us consider calculation of the average number of reflected and
transmitted photons. When $t \rightarrow \infty$, the average
numbers of reflected and transmitted photons are coupled by the
condition $\langle \hat{N}_{l}\rangle = n - \langle
\hat{N}_{r}\rangle$, where the average number of reflected photons
is given by
\begin{equation} \label{eq:b3}
  \langle \hat{N}_r\rangle = \frac{\Gamma}{2} \int ^{\infty}_0 \mathd \tau \, \langle n|\sigma_{+} \, \sigma_{-}|n\rangle_{\tau}.
\end{equation}
The integrand in (\ref{eq:b3}) is governed by the equation
\begin{equation} \label{eq:b4}
  \left(\partial_t + \Gamma\right)\langle n|\sigma_{+} \, \sigma_{-}|n\rangle =
  \mathi \, g \, \sqrt{n} \, A^{*}(t) \, \langle n-1|\sigma_{-}|n\rangle + \mathcc,
\end{equation}
which follows from Eq. (\ref{eq:eq17}). In turn, the matrix element $\langle n-1|\sigma_{-}|n\rangle$ obeys
\begin{equation} \label{eq:b5}
  \left(\partial_t + \mathi \, \Delta + \Gamma/2\right) \langle n-1|\sigma_{-}|n\rangle =
  2 \, \mathi \, g \, \sqrt{n} \, A(t) \, \langle n-1|\sigma_{+} \, \sigma_{-}|n-1\rangle - \mathi \, g \, \sqrt{n} \, A(t).
\end{equation}
To obtain Eqs. (\ref{eq:b4}) and (\ref{eq:b5}) we have used the relation
\begin{equation} \label{eq:b6}
  \int \mathd p \left[\tilde{l}_{p}(t) + \tilde{r}_{p}(t)\right]|n\rangle = \sqrt{n} \, A(t) \, |n-1\rangle.
\end{equation}

It can be seen that $\langle n|\sigma_{+} \, \sigma_{-}|n\rangle$
depends on $\langle n-1|\sigma_{-}|n\rangle$ which in turn depends
on $\langle n-1|\sigma_{+} \, \sigma_{-}|n-1\rangle$ and so on down
to $\langle 1|\sigma_{+} \, \sigma_{-}|1\rangle$ and $\langle
0|\sigma_{-}|1\rangle$. This set of $2n$ coupled equations should be
complemented by the initial conditions $\langle m|\sigma_{+} \,
\sigma_{-}|m\rangle|_{t=0} = \langle m-1|\sigma_{-}|m\rangle|_{t=0}
= 0$, where $1 \leq m \leq n$. For $n = 1$ we can use
(\ref{eq:eq28}) and write $\langle 1|\sigma_{+} \, \sigma_{-}|1\rangle =
\langle 1|\sigma_{+}|0\rangle\langle 0|\sigma_{-}|1\rangle =
\left|\langle 0|\sigma_{-}|1\rangle\right|^2$,
where $\langle 0|\sigma_{-}|1\rangle$ is given by (\ref{eq:eq29}).

The two-time correlation function $\langle n|\sigma_{+}(t) \, \sigma_{-}(t')|n\rangle$ is required for calculation of the phase-space distributions [see Eqs. (\ref{eq:eq23}) and (\ref{eq:eq25})].
The equation of motion for $\langle n|\sigma_{+}(t) \, \sigma_{-}(t')|n\rangle$ is given by
\begin{align} \label{eq:b7}
  \left(\partial_{t} - \mathi \, \Delta + \Gamma/2\right) \langle n|\sigma_{+}(t) \, \sigma_{-}(t')|n\rangle = &
  - 2 \, \mathi \, g \, \sqrt{n} \, A^{*}(t) \, \langle n-1|\sigma_{+}(t) \, \sigma_{-}(t) \, \sigma_{-}(t')|n\rangle \nonumber \\
  & + \mathi \, g \, \sqrt{n} \, A^{*}(t) \, \langle n-1|\sigma_{-}(t')|n\rangle,
\end{align}
where the initial value $\langle n|\sigma_{+}(t=t') \, \sigma_{-}(t')|n\rangle$ is taken from solution of Eqs. (\ref{eq:b4}) and (\ref{eq:b5}).
The evolution of $\langle n-1|\sigma_{+}(t) \, \sigma_{-}(t) \, \sigma_{-}(t')|n\rangle$ is governed by
\begin{align} \label{eq:b8}
  \left(\partial_{t} + \Gamma\right)\langle n-1|\sigma_{+}(t) \, \sigma_{-}(t) \, \sigma_{-}(t')|n\rangle =
  & \, \mathi \, g \, \sqrt{n-1} \, A^{*}(t) \, \langle n-2|\sigma_{-}(t) \, \sigma_{-}(t')|n\rangle \nonumber \\
  & - \mathi \, g \, \sqrt{n} \, A(t) \, \langle n - 1|\sigma_{+}(t) \, \sigma_{-}(t')|n-1\rangle.
\end{align}
The equation of motion for $\langle n - 1|\sigma_{+}(t) \, \sigma_{-}(t')|n-1\rangle$ entering the right side of (\ref{eq:b8}) is
\begin{align} \label{eq:b9}
  \left(\partial_{t} + \mathi \, \Delta + \Gamma/2\right) \langle n-2|\sigma_{-}(t) \, \sigma_{-}(t')|n\rangle =
  & \, 2 \, \mathi \, g \, \sqrt{n} \, A(t) \, \langle n-2|\sigma_{+}(t) \, \sigma_{-}(t) \, \sigma_{-}(t')|n-1\rangle \nonumber \\
  & - \mathi \, g \, \sqrt{n} \, A(t) \, \langle n-2|\sigma_{-}(t')|n-1\rangle.
\end{align}
We have used (\ref{eq:comm1}) to derive Eqs. (\ref{eq:b7}),
(\ref{eq:b8}), and (\ref{eq:b9}). Zero-value initial conditions (at
$t=t'$) should be imposed for solutions of Eqs. (\ref{eq:b8})
and (\ref{eq:b9}).
The equations (\ref{eq:b7})-(\ref{eq:b9}) show that $\langle n|\sigma_{+}(t) \, \sigma_{-}(t')|n\rangle$ can be expressed via
two-time functions with lower values of $n$ (down to $n=1$). For
$\langle 1|\sigma_{+}(t) \, \sigma_{-}(t')|1\rangle$ we can use the
explicit expression obtained with the use of (\ref{eq:eq28}) and
(\ref{eq:eq29}):
\begin{equation}
 \langle 1|\sigma_{+}(t) \, \sigma_{-}(t')|1\rangle = g^2 \int^{t}_{0} \mathd \tau \, \mathe^{(-\mathi \, \Delta + \Gamma/2)(t - \tau)} \, A(\tau) \int^{t'}_{0} \mathd \tau \, \mathe^{(-\mathi \, \Delta + \Gamma/2)(t' - \tau)} \, A(\tau).
\end{equation}
\end{document}